\documentclass[epj]{svjour}
\usepackage{graphics}

\usepackage{amssymb}
\usepackage{amsmath}
\usepackage{bm}
\usepackage{mathptmx}

\begin{document}
\title{Superconducting currents and charge gradients in the octonion spaces}
\author{Zi-Hua Weng\inst{1} \inst{2}
\thanks{\emph{Email address: } xmuwzh@xmu.edu.cn }%
}                     
%
%
\institute{School of Aerospace Engineering, Xiamen University, Xiamen 361005, China
\and College of Physical Science and Technology, Xiamen University, Xiamen 361005, China}
\date{Received: date / Revised version: date}
%
\abstract{
The paper focuses on applying the algebra of octonions to explore the influence of electric-charge gradients on the electric-current derivatives, revealing some of major influence factors of high pulse electric-currents. J. C. Maxwell was the first scholar to utilize the algebra of quaternions to study the physical properties of electromagnetic fields. The contemporary scholars employ simultaneously the quaternions and octonions to investigate the physical properties of electromagnetic fields, including the octonion field strength, field source, linear momentum, angular momentum, torque, and force and so forth. When the octonion force is equal to zero, it is able to achieve eight equations independent of each other, including the fluid continuity equation, current continuity equation, force equilibrium equation, and second-force equilibrium equation and so on. One of inferences derived from the second-force equilibrium equation is that the charge gradient and current derivative are interrelated closely, two of them must satisfy the need of the second-force equilibrium equation synchronously. Meanwhile the electromagnetic strength and linear momentum both may exert an influence on the current derivative to a certain extent. The above states that the charge gradient and current derivative are two correlative physical quantities, they must meet the requirement of second-force equilibrium equation. By means of controlling the charge gradients and other physical quantities, it is capable of restricting the development process of current derivatives, reducing the damage caused by the instantaneous impact of high pulse electric-currents, enhancing the anti-interference ability of electronic equipments to resist the high pulse electric-currents and their current derivatives. Further the second-force equilibrium equation is able to explain two types of superconducting currents.
} 
\maketitle

\section{\label{sec:level1}Introduction}

The steep and powerful electric-current derivatives may produce the significant impact and damage to the electronic components, electronic circuits, and high-voltage transmission lines and others. Can we find some appropriate methods to reduce actively the impact and damage of current derivatives? What are the influencing factors of current derivatives?  For many years, these problems have puzzled scholars and engineers.  They introduce a variety of detection methods to measure the steep current derivatives, attempting to decrease the harm of current derivatives. It was not until the emergence of the field theory, described with the octonions, that these questions were answered to a certain extent. The field theory reveals that the current derivatives and electric-charge gradient are interrelated rather than uncorrelated. By means of controlling the charge gradient, the amplitude of current derivative can be reduced effectively, decreasing the harm extent of current derivatives, enhancing the ability of electronic equipments and facilities to resist the impact of current derivatives.

In the theoretical research of electromagnetic fields, the electric currents and their derivatives play an important role. The current derivative has produced a lot of impact and damage to the electronic facilities. Especially, the emerged high pulse electric-current and its current derivative may destroy suddenly some expensive electronic components, electronic circuits, and switchgears for the high-voltage transmission lines and others. The related technology of the high pulse electric-current and its current derivative can be applied in some research areas, for instance, the electromagnetic pulse technology \cite{lekner,fang,borhanian}, controlled fusion \cite{wu}, thermal plasma \cite{zhang}, pulse laser technology \cite{zhao}, and high voltage process \cite{garcia} and others.

In the practical engineering application, the scholars and engineers have spent lots of time and endeavor on the item, introducing a variety of methods to resist the threat of the high pulse electric-current and its current derivative. The application scope of the current derivative is very expansive, mainly involving the electronic control system \cite{idris}, direct-current (DC for short) electromotor \cite{mokin}, DC traction power supply system \cite{yang}, quick short-circuit protection circuits \cite{saeed}, DC power changing set of silicon controlled rectifier \cite{chen}, automatic speed regulator \cite{bansal}, reversible speed control system \cite{qczhang}, current-limiting fuse \cite{torres}, vacuum evaporation equipment \cite{siniavskii}, CMOS and IGBT Modules \cite{gorecki,lu} and others.

Up till the present moment, we can only passively detect the abruptly appearance of high pulse electric-currents and their current derivatives in the majority of cases, taking measures to solve these relevant problems immediately. The experts should analyze the internal causes of the sudden appearance of high pulse electric-currents and relevant current derivatives, taking necessary measures to actively suppress the harm of high pulse electric-currents. Consequently, it is necessary to focus on some rules that must be observed in the study of the high pulse electric-currents and their current derivatives.

According to the above analysis and comparison, there are some shortcomings in the study of current derivatives and charge gradients, in the classical electromagnetic theory. It restricts the further exploration and application of the current derivatives and charge gradients.

1) Current derivative. In the past, the current derivative was considered as an isolated physical quantity. And it was merely related to the electric current and time, but not to other physical quantities. Especially, the current derivative and charge gradient were regarded as two independent physical quantities. This recognition is not conducive to the suppression of current derivatives from the perspective of internal factors, and it is not beneficial to reduce the impact and threat of current derivatives in essence.

2) Electromagnetic strength. The electromagnetic strength and gravitational strength both were considered to have no effect on the current derivatives or charge gradients. The possible correlation between the electromagnetic strength and current derivative is also often mistaken for interference. The electromagnetic strength and current derivative are even supposed to be uncorrelated. Furthermore, the gravitational strength was preset to be unable to make a contribution to the current derivative or charge gradient either.

3) Spatial position. The relative position of the linear momentum and magnetic induction intensity in the space was not considered to be able to make a contribution to the current derivative or charge gradient. The linear momentum and magnetic induction intensity were conceived to be independent of each other, while their relative spatial positions were believed to be impossible to exert an impact on other physical quantities. The latent correlation between the linear momentum and current derivative was ignored undoubtedly.

By all appearances, the classical electromagnetic theory cannot effectively study the relationship between the current derivative and charge gradient. Either it is unable to reveal the influence of electromagnetic strength, gravitational strength and relative spatial position on the current derivatives. This limitation will restrict the application scope of the existing electromagnetic theory.

In a sharp contrast to the above, the electromagnetic and gravitational theory, described with the algebra of octonions (octonion field theory, for short temporarily), is capable of unpuzzling the influence of charge gradients and others on the current derivatives. It reveals that the charge gradient, electromagnetic strength, gravitational strength, and spatial position and others will make a contribution to the current derivatives. Some inferences derived from the octonion field theory extend the application scope of electromagnetic theory.

J. C. Maxwell first applied the algebra of quaternions to study the physical properties of electromagnetic fields. This method inspired the subsequent scholars to use the quaternions to research the physical properties of electromagnetic fields and gravitational fields \cite{mironov1}. Further the scholars utilize the algebra of octonions to investigate the quantum effect \cite{deleo1,gogberashvili,bernevig}, wave equation \cite{deleo2,deleo3,deleo4}, relativity \cite{bossard}, curved space, dark matter \cite{furui}, dark energy, weak nuclear fields \cite{majid,farrill}, strong nuclear fields \cite{chanyal1,furey}, continuity equations \cite{tanisli1} and equilibrium equations and others. Especially, it is able to employ the algebra of octonions to explore simultaneously the physical properties of electromagnetic and gravitational fields, including the octonion field potential, field strength, field source, linear momentum, angular momentum, torque and force and so forth.

The application of octonions is able to describe simultaneously the theories of electromagnetic fields and gravitational fields. When the octonion force (in Section 3) is equal to zero, it is capable of achieving eight equations independent of each other. One of the eight independent equations reveals that the current derivative is interrelated with the charge gradient and others. The charge gradient (in Section 4) can directly exert a significant impact on the current derivative.

In the octonion field theory, the physical properties of the current derivatives and charge gradients can be effectively studied by the application of the algebra of octonions. This method possesses some important advantages as follows.

1) Charge gradient. The current derivative and charge gradient are two interrelated physical quantities, but the second-force equilibrium equation must be satisfied between the two of them (see Section 3). The variation of charge gradient will induce the fluctuation of current derivative and vice versa. The current derivative and charge gradient, in the second-force equilibrium equation, are similar to the inertial force and energy gradient respectively, in the force equilibrium equation.

2) Field strength. Both of gravitational strength and electromagnetic strength are able to make a contribution to the second-force equilibrium equation to a certain extent. By means of the application of the electromagnetic strength, we can dominate the alteration of the current derivative and charge gradient. Especially, the ultra-strong magnetic induction intensity or electric field intensity are capable of impacting the current derivative and charge gradient evidently.

3) Relative position. In the relative spatial position, there are certain interactions and correlations among the physical quantities of gravitational fields with that of electromagnetic fields. The relative spatial position, between the linear momentum and magnetic induction intensity, will exert an impact on the current derivative and ultra-strong charge gradient. Similarly, the relative spatial position, between the electric current and ultra-strong gravitational precession-angular-velocity (in Section 2), may also have an influence on the second-force equilibrium equation to a certain extent.

Making use of the octonion field theory, the paper explores the cause of current derivative and relevant influencing factors, from the point of view of the relationship between the current derivative and charge gradient. It is revealed that neither the current derivative nor the charge gradient is one isolated physical quantity. By means of the modulation of the charge gradient, it is able to suppress the amplitude of current derivative, depressing the impact on the electronic equipments, enhancing the ability of electronic equipments to resist the current derivatives of the high pulse electric-currents.

J. T. Graves and A. Cayley independently introduced the octonion, which is called as the standard octonion. What the paper discusses is the standard octonion, rather than the non-standard octonion. The latter includes the split-octonions \cite{chanyal2,tanisli2,mironov2}, Cartan's octonions \cite{baez}, and pseudo-octonions and others. Each non-standard octonion can be considered as one function of standard octonion, in the mathematics. The standard and non-standard octonions are able to describe different types of physical properties. Consequently the standard octonions and non-standard octonions can be combined together to become a whole, describing jointly various physical properties.

The standard octonions can be applied to explore the optics, computer graphics, satellite maneuver control, and camera attitude control and others, beside the physical theories and mathematics. For example, as one part of standard octonions, the standard quaternions can express effectively the Eulerian angles, including the nutation angle, precession angle, and intrinsic rotation angle. If one utilizes the vector terminology to depict the Eulerian angles of satellite orbits, each matrix element of a few crucial matrixes may be simultaneously equal to zero occasionally. But it never happens when one applies the standard quaternions to express the Eulerian angles of satellite orbits. By comparison with the vector terminology, the standard quaternions possess obvious advantages to describe the Eulerian angles.

By virtue of the algebra of octonions, it is able to describe the physical properties of electromagnetic and gravitational fields.

\begin{table}[h]
\centering
\caption{The multiplication table of standard octonions, which were introduced by J. T. Graves and A. Cayley independently.}
\label{tab:table1}
\begin{tabular}{ccccccccc}
\hline\noalign{\smallskip}
$ $            & $1$            & $\textbf{d}_1$  & $\textbf{d}_2$  & $\textbf{d}_3$  & $\textbf{D}_0$  & $\textbf{D}_1$ & $\textbf{D}_2$  & $\textbf{D}_3$    \\
\noalign{\smallskip}\hline\noalign{\smallskip}
$1$            & $1$            & $\textbf{d}_1$  & $\textbf{d}_2$  & $\textbf{d}_3$  & $\textbf{D}_0$  & $\textbf{D}_1$  & $\textbf{D}_2$  & $\textbf{D}_3$   \\
$\textbf{d}_1$ & $\textbf{d}_1$ & $-1$            & $\textbf{d}_3$  & $-\textbf{d}_2$ & $\textbf{D}_1$  & $-\textbf{D}_0$ & $-\textbf{D}_3$ & $\textbf{D}_2$   \\
$\textbf{d}_2$ & $\textbf{d}_2$ & $-\textbf{d}_3$ & $-1$            & $\textbf{d}_1$  & $\textbf{D}_2$  & $\textbf{D}_3$  & $-\textbf{D}_0$ & $-\textbf{D}_1$  \\
$\textbf{d}_3$ & $\textbf{d}_3$ & $\textbf{d}_2$  & $-\textbf{d}_1$ & $-1$            & $\textbf{D}_3$  & $-\textbf{D}_2$ & $\textbf{D}_1$  & $-\textbf{D}_0$  \\
\noalign{\smallskip}\hline\noalign{\smallskip}
$\textbf{D}_0$ & $\textbf{D}_0$ & $-\textbf{D}_1$ & $-\textbf{D}_2$ & $-\textbf{D}_3$ & $-1$            & $\textbf{d}_1$  & $\textbf{d}_2$  & $\textbf{d}_3$   \\
$\textbf{D}_1$ & $\textbf{D}_1$ & $\textbf{D}_0$  & $-\textbf{D}_3$ & $\textbf{D}_2$  & $-\textbf{d}_1$ & $-1$            & $-\textbf{d}_3$ & $\textbf{d}_2$   \\
$\textbf{D}_2$ & $\textbf{D}_2$ & $\textbf{D}_3$  & $\textbf{D}_0$  & $-\textbf{D}_1$ & $-\textbf{d}_2$ & $\textbf{d}_3$  & $-1$            & $-\textbf{d}_1$  \\
$\textbf{D}_3$ & $\textbf{D}_3$ & $-\textbf{D}_2$ & $\textbf{D}_1$  & $\textbf{D}_0$  & $-\textbf{d}_3$ & $-\textbf{d}_2$ & $\textbf{d}_1$  & $-1$             \\
\noalign{\smallskip}\hline
\end{tabular}
\end{table}

\section{Octonion spaces}

In the octonion field theory, the quaternion space can be applied to describe the physical properties of either gravitational fields or electromagnetic fields. The quaternion space $\mathbb{H}_g$ is able to depict the physical properties of gravitational fields, while the 2-quaternion space (short for the second quaternion space) $\mathbb{H}_e$ is capable of picturing that of electromagnetic fields. The two quaternion spaces, $\mathbb{H}_g$ and $\mathbb{H}_e$ , are independent of each other. Apparently, they can be combined together to become an octonion space $\mathbb{O}$.

\subsection{Octonion field source}

In the quaternion space $\mathbb{H}_g$ for gravitational fields, the radius vector is, $\mathbb{R}_g = i r_0 \textbf{d}_0 + \Sigma r_k \textbf{d}_k$ , and the velocity is, $\mathbb{V}_g = i v_0 \textbf{d}_0 + \Sigma v_k \textbf{d}_k$. The gravitational potential is, $\mathbb{A}_g = i a_0 \textbf{d}_0 + \Sigma a_k \textbf{d}_k$ , the gravitational strength is, $\mathbb{F}_g = f_0 \textbf{d}_0 + \Sigma f_k \textbf{d}_k$, and the gravitational source is, $\mathbb{S}_g = i s_0 \textbf{d}_0 + \Sigma s_k \textbf{d}_k$ . Herein $r_j$ , $v_j$ , $a_j$ , $s_j$ , and $f_0$ are all real. $f_k$ is one complex number. $i$ is the imaginary unit. $v_0$ is the speed of light. $r_0 = v_0 t$ , and $t$ is the time. $\textbf{a} = \Sigma a_k \textbf{d}_k$. $\textbf{d}_j$ is the basis vector. $\textbf{d}_0 = 1$. $\textbf{d}_k^2 = - 1$. $j = 0, 1, 2, 3$. $k = 1, 2, 3$.

Similarly, in the 2-quaternion space $\mathbb{H}_e$ for electromagnetic fields, the radius vector is, $\mathbb{R}_e = i R_0 \textbf{D}_0 + \Sigma R_k \textbf{D}_k$ , and the velocity is, $\mathbb{V}_e = i V_0 \textbf{D}_0 + \Sigma V_k \textbf{D}_k$ . The electromagnetic potential is, $\mathbb{A}_e = i A_0 \textbf{D}_0 + \Sigma A_k \textbf{D}_k$ , the electromagnetic strength is, $\mathbb{F}_e = F_0 \textbf{D}_0 + \Sigma F_k \textbf{D}_k$, and the electromagnetic source is, $\mathbb{S}_e = i S_0 \textbf{D}_0 + \Sigma S_k \textbf{D}_k$ . Herein $R_j$ , $V_j$ , $A_j$ , $S_j$, and $F_0$ are all real. $F_k$ is one complex number. $\textbf{D}_j$ is the basis vector. $\textbf{D}_j^2 = - 1$. $\textbf{A}_0 = A_0 \textbf{D}_0$, and $\textbf{A} = \Sigma A_k \textbf{D}_k$ . The selection of coordinate axes \cite{weng1} in the quaternion space $\mathbb{H}_e$ is similar to that in the quaternion space $\mathbb{H}_g$ .

The two independent quaternion spaces, $\mathbb{H}_g$ and $\mathbb{H}_e$ , can be considered to be perpendicular to each other. As a result, the two quaternion spaces will be combined together to become one octonion space, $\mathbb{O} = \mathbb{H}_g + \mathbb{H}_e$ , with $\textbf{D}_j = \textbf{d}_j \circ \textbf{D}_0$ . The octonion space $\mathbb{O}$ is able to explore simultaneously the physical properties of gravitational and electromagnetic fields (Table \ref{tab:table1}). In the octonion space $\mathbb{O}$ for gravitational and electromagnetic fields, the octonion field potential is, $\mathbb{A} = \mathbb{A}_g + k_{eg} \mathbb{A}_e$ , the octonion field strength is, $\mathbb{F} = \mathbb{F}_g + k_{eg} \mathbb{F}_e$ , and the octonion field source is, $\mathbb{S} = \mathbb{S}_g + k_{eg} \mathbb{S}_e$ . Herein $k_{eg}$ is one coefficient, meeting the demand of the dimensional homogeneity \cite{weng2}.

From the octonion field potential $\mathbb{A}$ , it is able to define the octonion field strength as follows (Table \ref{tab:table2}),
\begin{eqnarray}
\mathbb{F} = \lozenge \circ \mathbb{A} ~,
\end{eqnarray}
where it is able to select $f_0 = 0$ as the gauge condition in the gravitational fields, so the gravitational strength is, $\textbf{f} = \Sigma f_k \textbf{d}_k$ , with $\textbf{f} = i \textbf{g} / v_0 + \textbf{b}$ . $\textbf{g}$ is the gravitational acceleration, while $\textbf{b}$ is the gravitational precession-angular-velocity. Similarly, $F_0 = 0$ can be chosen as the gauge condition in the electromagnetic fields, therefore the electromagnetic strength will be simplified into, $\textbf{F} = \Sigma F_k \textbf{D}_k $, with $\textbf{F} = i \textbf{E} / v_0 + \textbf{B}$ . $\textbf{E}$ is the electric field intensity, and $\textbf{B}$ is the magnetic induction intensity. $\circ$ denotes the octonion multiplication. $\lozenge = \emph{i} \partial_0 + \Sigma \emph{\textbf{d}}_k \partial_k$ . $ \nabla = \Sigma \emph{\textbf{d}}_k \partial_k$ , $ \partial_j = \partial / \partial r_j$ .

Furthermore, from the octonion field strength $\mathbb{F}$ , one can define the octonion field source (Table \ref{tab:table3}),
\begin{eqnarray}
\mu \mathbb{S} && = - ( i \mathbb{F} / v_0 + \lozenge )^\ast \circ \mathbb{F}
\nonumber
\\
&&
= \mu_g \mathbb{S}_g + k_{eg} \mu_e \mathbb{S}_e - i \mathbb{F}^\ast \circ \mathbb{F} / v_0  ~,
\end{eqnarray}
where $\mu$ is one coefficient. $\mu_g$ is the gravitational constant, while $\mu_e$ is the electromagnetic constant. $\ast$ indicates the octonion conjugate.

\begin{table}[h]
\centering
\caption{The multiplication of two operators, $\nabla$ and $\partial_0$ , with the octonion physics quantities.}
\label{tab:table2}
\begin{tabular}{ll}
\hline\noalign{\smallskip}
definition                         &     expression~meaning                                                          \\
\noalign{\smallskip}\hline\noalign{\smallskip}
$\nabla \cdot \textbf{a}$          &     $-(\partial_1 a_1 + \partial_2 a_2 + \partial_3 a_3)$                       \\
$\nabla \times \textbf{a}$         &     $\textbf{d}_1 ( \partial_2 a_3 - \partial_3 a_2 )
                                           + \textbf{d}_2 ( \partial_3 a_1 - \partial_1 a_3 )
                                           + \textbf{d}_3 ( \partial_1 a_2 - \partial_2 a_1 )$                       \\
$\nabla a_0$                       &     $\textbf{d}_1 \partial_1 a_0
                                           + \textbf{d}_2 \partial_2 a_0
                                           + \textbf{d}_3 \partial_3 a_0  $                                          \\
$\partial_0 \textbf{a}$            &     $\textbf{d}_1 \partial_0 a_1
                                           + \textbf{d}_2 \partial_0 a_2
                                           + \textbf{d}_3 \partial_0 a_3  $                                          \\
\noalign{\smallskip}\hline\noalign{\smallskip}
$\nabla \cdot \textbf{A}$          &     $-(\partial_1 A_1 + \partial_2 A_2 + \partial_3 A_3) \textbf{D}_0 $         \\
$\nabla \times \textbf{A}$         &     $-\textbf{D}_1 ( \partial_2 A_3 - \partial_3 A_2 )
                                           - \textbf{D}_2 ( \partial_3 A_1 - \partial_1 A_3 )
                                           - \textbf{D}_3 ( \partial_1 A_2 - \partial_2 A_1 )$                       \\
$\nabla \circ \textbf{A}_0$        &     $\textbf{D}_1 \partial_1 A_0
                                           + \textbf{D}_2 \partial_2 A_0
                                           + \textbf{D}_3 \partial_3 A_0  $                                          \\
$\partial_0 \textbf{A}$            &     $\textbf{D}_1 \partial_0 A_1
                                           + \textbf{D}_2 \partial_0 A_2
                                           + \textbf{D}_3 \partial_0 A_3  $                                          \\
\noalign{\smallskip}\hline
\end{tabular}
\end{table}

\subsection{Octonion force}

From the octonion field source, it is capable of defining the octonion linear momentum (Table \ref{tab:table4}),
\begin{eqnarray}
\mathbb{P} = \mu \mathbb{S} / \mu_g ~,
\end{eqnarray}
where $\mathbb{P} = \mathbb{P}_g + k_{eg} \mathbb{P}_e$ . $\mathbb{P}_g = \{ \mu_g \mathbb{S}_g - i \mathbb{F}^\ast \circ \mathbb{F} / v_0 \} / \mu_g$ , $\mathbb{P}_e = \mu_e \mathbb{S}_e / \mu_g$ . $\mathbb{P}_g = i p_0 + \textbf{p}$ . $\mathbb{P}_e = i \textbf{P}_0 + \textbf{P}$ . $\textbf{p} = \Sigma p_k \textbf{d}_k$ , $\textbf{P} = \Sigma P_k \textbf{D}_k$ , $\textbf{P}_0 = P_0 \textbf{D}_0$ . $p_j$ and $P_j$ are all real.

The octonion angular momentum is defined as,
\begin{eqnarray}
\mathbb{L} = ( \mathbb{R} + k_{rx} \mathbb{X} )^\times \circ \mathbb{P} ~,
\end{eqnarray}
where $\mathbb{R} = \mathbb{R}_g + k_{eg} \mathbb{R}_e$ . $\mathbb{X}$ is the octonion integrating function of field potential (see Ref.\cite{weng1}). $\times$ is the complex conjugate.

The octonion torque $\mathbb{W}$ can be defined from the octonion angular momentum $\mathbb{L}$ ,
\begin{eqnarray}
\mathbb{W} = - v_0 ( i \mathbb{F} / v_0 + \lozenge ) \circ \mathbb{L} ~,
\end{eqnarray}
and the octonion force $\mathbb{N}$ may be defined from the octonion $\mathbb{W}$ ,
\begin{eqnarray}
\mathbb{N} = - ( i \mathbb{F} / v_0 + \lozenge ) \circ \mathbb{W} ~,
\end{eqnarray}
where $\mathbb{N} = \mathbb{N}_g + k_{eg} \mathbb{N}_e$ . $\mathbb{N}_g = i N_{10}^i + N_{10} + i \textbf{N}_1^i + \textbf{N}_1$ . $\mathbb{N}_e = i \textbf{N}_{20}^i + \textbf{N}_{20} + i \textbf{N}_2^i + \textbf{N}_2$ .

In the octonion space $\mathbb{O}$ , in case the octonion force equals to zero, $\mathbb{N} = 0$, it is able to achieve,
\begin{eqnarray}
i N_{10}^i + N_{10} + i \textbf{N}_1^i + \textbf{N}_1 = 0 ~,
\\
i \textbf{N}_{20}^i + \textbf{N}_{20} + i \textbf{N}_2^i + \textbf{N}_2 = 0 ~,
\end{eqnarray}
further we can deduce eight equilibrium equations and continuity equations from the above.

\begin{table}[h]
\centering
\caption{Some definitions and physics quantities in the gravitational and electromagnetic fields.}
\label{tab:table3}
\begin{tabular}{lll}
\hline\noalign{\smallskip}
physics~quantity             &      definition                                                                                   \\
\noalign{\smallskip}\hline\noalign{\smallskip}
quaternion~operator          &      $\lozenge = i \partial_0 + \Sigma \emph{\textbf{d}}_k \partial_k$                            \\
radius~vector                &      $\mathbb{R} = \mathbb{R}_g + k_{eg} \mathbb{R}_e  $                                          \\
integral~function            &      $\mathbb{X} = \mathbb{X}_g + k_{eg} \mathbb{X}_e  $                                          \\
field~potential              &      $\mathbb{A} = i \lozenge^\times \circ \mathbb{X}  $                                          \\
field~strength               &      $\mathbb{F} = \lozenge \circ \mathbb{A}  $                                                   \\
field~source                 &      $\mu \mathbb{S} = - ( i \mathbb{F} / v_0 + \lozenge )^* \circ \mathbb{F} $                   \\
linear~momentum              &      $\mathbb{P} = \mu \mathbb{S} / \mu_g $                                                       \\
angular~momentum             &      $\mathbb{L} = ( \mathbb{R} + k_{rx} \mathbb{X} )^\times \circ \mathbb{P} $                   \\
octonion~torque              &      $\mathbb{W} = - v_0 ( i \mathbb{F} / v_0 + \lozenge ) \circ \mathbb{L} $                     \\
octonion~force               &      $\mathbb{N} = - ( i \mathbb{F} / v_0 + \lozenge ) \circ \mathbb{W} $                         \\
\noalign{\smallskip}\hline
\end{tabular}
\end{table}

\section{Equilibrium equations}

When the octonion force is equal to zero, it is able to deduce eight equations from $\mathbb{N} = 0$, that is, four equations in the quaternion space $\mathbb{H}_g$ for gravitational fields, and four equations in the 2-quaternion space $\mathbb{H}_e$ for electromagnetic fields. Up to now, we have found the physical meaning of six out of the eight equations, including the second-force equilibrium equation discussed in the paper.

\subsection{Quaternion space}

In the quaternion space $\mathbb{H}_g$ , from Eq.(7), it is capable of inferring four equations simultaneously, that is, $N_{10}^i = 0$, $N_{10} = 0$, $\textbf{N}_1^i = 0$, and $\textbf{N}_1 = 0$. By means of expanding the four equations respectively, we can find that two of them are equilibrium equations, and the rest are equilibrium equations.

It is easy to find that the two of four equations can be respectively degenerated into the fluid continuity equation and force equilibrium equation in the classical theory, under some approximate cases. Further we found the physical significance of the rest of four equations. Nowadays the physical meanings of all of four equations are explicit in the quaternion space $\mathbb{H}_g$ . They are respectively the force equilibrium equation, fluid continuity equation, precession equilibrium equation, and torque continuity equation.

1) Force equilibrium equation. The inertial force, that the objects suffered, and applied forces (such as, energy gradient, gravitational force, and electromagnetic force) and others must obey the force equilibrium equation, $\textbf{N}_1^i = 0$.

2) Fluid continuity equation. The mass and linear momentum and others in the fluids have to meet the need of the fluid continuity equation, $N_{10} = 0$.

3) Precession equilibrium equation. The precessional angular velocity of objects and applied torque and others will meet the requirements of the precession equilibrium equation, $\textbf{N}_1 = 0$. The precessional angular velocity and applied torque derivative, in the precession equilibrium equation, are respectively similar to the acceleration and applied force, in the force equilibrium equation.

4) Torque continuity equation. The divergence of angular momentum and applied torque and others in the fluids can fulfill a requirement of the torque continuity equation, $N_{10}^i = 0$. The divergence of angular momentum and applied torque, in the torque continuity equation, are respectively analogous to the mass and linear momentum, in the fluid continuity equation.

\subsection{Second-quaternion space}

In the 2-quaternion space $\mathbb{H}_e$ , from Eq.(8), one can simultaneously deduce four equations, that is, $\textbf{N}_{20}^i = 0$, $\textbf{N}_{20} = 0$, $\textbf{N}_2^i = 0$, and $\textbf{N}_2 = 0$. Making use of expanding the four equations respectively, it is found that two of them belong to the equilibrium equations, while the rest are owned by the equilibrium equations.

It is interesting to note that each of the four equations is the vector equation, in the 2-quaternion space $\mathbb{H}_e$. As a result, the continuity equation is essentially consistent with the equilibrium equation, according to the point of view of the octonion field theory in the octonion space.

Compared with the classical theory, the four equations cover the current continuity equation, under the approximate cases. Until now it is found the physical meaning of two out of four equations, in the 2-quaternion space $\mathbb{H}_e$ . They are respectively the current continuity equation and second-force equilibrium equation.

1) Current continuity equation. The electric charge and electric current and others satisfy a requirement of the current continuity equation, $\textbf{N}_{20} = 0$. Obviously, the electric charge and electric current, in the current continuity equation, are respectively akin to the mass and linear momentum, in the fluid continuity equation.

2) Second-force equilibrium equation. The current derivative and charge gradient and others meet a requirement of the equation, $\textbf{N}_2^i = 0$, which is called as the second-force equilibrium equation temporarily. The current derivative and charge gradient, in the second-force equilibrium equation, resemble respectively the inertial force and energy gradient, in the force equilibrium equation (Table \ref{tab:table5}).

The paper aims to explore the contribution of the current derivatives and charge gradients and others on the second-force equilibrium equation.

\begin{table}[h]
\caption{Comparison of physics quantities between the gravitational fields with electromagnetic fields, in the complex octonion spaces.}
\centering
\label{tab:table4}
\begin{tabular}{lll}
\hline\noalign{\smallskip}
physics~quantity                   &  gravitational~field                                         &  electromagnetic~field                                          \\
\noalign{\smallskip}\hline\noalign{\smallskip}
field~potential, $\mathbb{A}$      &  $a_0$, gravitational~scalar~potential                       &  $\textbf{A}_0$, electromagnetic~scalar~potential               \\
                                   &  $\textbf{a}$, gravitational~vector~potential                &  $\textbf{A}$, electromagnetic~vector~potential                 \\
field~strength, $\mathbb{F}$ 	   &  $\textbf{g}$, gravitational~acceleration 	                  &  $\textbf{E}$, electric~field~intensity                         \\
                  	               &  $\textbf{b}$, gravitational precession-angular-velocity     &  $\textbf{B}$, magnetic~induction~intensity                     \\
field~source, $\mathbb{S}$ 	       &  $s_0$, mass~density 	                                      &  $\textbf{S}_0$, electric-charge~density                        \\
         	                       &  $\textbf{s}$, linear-momentum~density 	                  &  $\textbf{S}$, electric-current~density                         \\
angular~momentum, $\mathbb{L}$ 	   &  $L_{10}$, dot~product                                       &  $\textbf{L}_{20}$, (similar to $L_{10}$)                       \\
	                               &  $\textbf{L}_1$, angular~momentum 	                          &  $\textbf{L}_2$, magnetic~moment                                \\
	                               &  $\textbf{L}_1^i$, (similar to $\textbf{L}_2^i$)             &  $\textbf{L}_2^i$, electric~moment                              \\
octonion~torque, $\mathbb{W}$ 	   &  $W_{10}$, divergence~of~angular~momentum                    &  $\textbf{W}_{20}$, divergence~of~magnetic~moment               \\
	                               &  $W_{10}^i$, energy                                          &  $\textbf{W}_{20}^i$, second-energy (similar to $W_{10}^i$)     \\
	                               &  $\textbf{W}_1$, curl~of~angular~momentum                    &  $\textbf{W}_2$, curl~of~magnetic~moment                        \\
	                               &  $\textbf{W}_1^i$, torque                                    &  $\textbf{W}_2^i$, second-torque (similar to $\textbf{W}_1^i$)  \\
octonion~force, $\mathbb{N}$       &  $N_{10}$, power 	                                          &  $\textbf{N}_{20}$, second-power (similar to $N_{10}$)          \\
	                               &  $N_{10}^i$, torque~divergence                               &  $\textbf{N}_{20}^i$, (similar to $N_{10}^i$)                   \\
	                               &  $\textbf{N}_1$, torque~derivative                           &  $\textbf{N}_2$, (similar to $\textbf{N}_1$)                    \\
	                               &  $\textbf{N}_1^i$, force                                     &  $\textbf{N}_2^i$, second-force (similar to $\textbf{N}_1^i$)   \\
\noalign{\smallskip}\hline
\end{tabular}
\end{table}

\section{Second-force equilibrium equation}

In the 2-quaternion space $\mathbb{H}_e$ , the current derivatives and charge gradients must satisfy the requirement of the second-force equilibrium equation. Meanwhile, it is also found that the electromagnetic strength and gravitational strength and others may exert an impact on the second-force equilibrium equation.

Further, we can expand the second-force equilibrium equation, $\textbf{N}_2^i = 0$. As a result, it can be written as follows (see Ref.\cite{weng1}),
\begin{eqnarray}
0 = && ( \textbf{g} \circ \textbf{W}_{20}^i / v_0 + \textbf{g} \times \textbf{W}_2^i / v_0 - \textbf{b} \circ \textbf{W}_{20} - \textbf{b} \times \textbf{W}_2 ) / v_0
\nonumber
\\
&& + ( W_{10}^i \textbf{E} / v_0 + \textbf{E} \times \textbf{W}_1^i / v_0 - W_{10} \textbf{B} - \textbf{B} \times \textbf{W}_1 ) / v_0
\nonumber
\\
&& - ( \partial_0 \textbf{W}_2 + \nabla \circ \textbf{W}_{20}^i + \nabla \times \textbf{W}_2^i ) ~ ,
\end{eqnarray}
where $k$ is the spatial dimension of the radius vector $\textbf{r}$ . $k_p = k - 1$. $\textbf{P} = ( \mu_e / \mu_g ) \textbf{S}$ . $\textbf{W}_2 \approx v_0 k_p \textbf{P}$ . $\textbf{W}_{20}^i \approx v_0 k_p \textbf{P}_0$ . $W_{10}^i \approx v_0 k_p p_0$. $\textbf{W}_1 \approx v_0 k_p \textbf{p}$ .

The above can be reduced into,
\begin{eqnarray}
0 = ( \textbf{g} \circ \textbf{W}_{20}^i / v_0 - \textbf{b} \circ \textbf{W}_{20} + W_{10}^i \textbf{E} / v_0 - W_{10} \textbf{B} ) / v_0 - \partial_0 \textbf{W}_2 ~ ,
\end{eqnarray}
and it means that the electromagnetic strength and gravitational strength and others will exert a significant impact on the derivative of magnetic moment or the current derivative.

Sometimes the derivatives, $\partial_t \textbf{B}$ , $\partial_t \textbf{E}$ , $\partial_t \textbf{g}$ , and $\partial_t \textbf{b}$ and so forth are tiny enough, and they can be neglected. So the second-force equilibrium equation, Eq.(9), can be approximately simplified as,
\begin{eqnarray}
0 = ( \textbf{g} \circ \textbf{P}_0 - v_0 \textbf{b} \times \textbf{P} ) / v_0 + ( p_0 \textbf{E} - v_0 \textbf{B} \times \textbf{p} ) / v_0 - v_0 ( \partial_0 \textbf{P} + \nabla \circ \textbf{P}_0 ) ~,
\nonumber
\end{eqnarray}
or
\begin{eqnarray}
0 = ( \partial_t \textbf{S} + v_0 \nabla \circ \textbf{S}_0 ) - ( \textbf{g} \circ \textbf{S}_0 / v_0 - \textbf{b} \times \textbf{S} ) - ( \mu_g / \mu_e ) ( p_0 \textbf{E} / v_0 - \textbf{B} \times \textbf{p} ) ~,
\end{eqnarray}
where $\partial_t \textbf{S}$ is the derivative of electric current, while $\nabla \circ \textbf{S}_0 / v_0$ is the gradient of electric charge. $\textbf{S}$ is the density of electric current, and $q$ is the density of electric charge. $\textbf{S}_0 = S_0 \textbf{D}_0$ . $S_0 = q v_0$ . $\partial_t = \partial / \partial t$ .

1) When the electromagnetic strength $( \textbf{E} , \textbf{B} )$ is comparatively weak, while the gravitational strength $( \textbf{g} , \textbf{b} )$ is comparatively strong, and the coefficient $( \mu_g / \mu_e )$ is quite small, Eq.(11) can be degenerated to,
\begin{eqnarray}
0 = ( \partial_t \textbf{S} + v_0 \nabla \circ \textbf{S}_0 ) - ( \textbf{g} \circ \textbf{S}_0 / v_0 - \textbf{b} \times \textbf{S} )  ~.
\end{eqnarray}

The above states that the comparatively strong gravitational strength will make a contribution to the second-force equilibrium equation. Especially, the vector product of the electric current with the gravitational precession-angular-velocity may impact directly the current derivatives or charge gradients.

2) When the gravitational strength $( \textbf{g} , \textbf{b} )$ is comparatively weak, while the electromagnetic strength $( \textbf{E} , \textbf{B} )$ is comparatively strong, Eq.(11) can be reduced to,
\begin{eqnarray}
0 = ( \partial_t \textbf{S} + v_0 \nabla \circ \textbf{S}_0 ) - ( \mu_g / \mu_e ) ( p_0 \textbf{E} / v_0 - \textbf{B} \times \textbf{p} ) ~.
\end{eqnarray}

The above means that the comparatively strong electromagnetic strength may exert an influence on the second-force equilibrium equation. Especially, the vector product of the linear momentum with the magnetic induction intensity will affect directly the current derivatives or charge gradients.

3) When the gravitational strength and electromagnetic strength both are comparatively weak, Eq.(11) can be simplified into,
\begin{eqnarray}
0 =  \partial_t \textbf{S} + v_0 \nabla \circ \textbf{S}_0  ~.
\end{eqnarray}

The above shows that the current derivative and charge gradient are closely correlated, in the case of comparatively weak field strength. And the second-force equilibrium equation should be satisfied between two of them.

In some materials, when the charge gradient is, $v_0 \nabla \circ \textbf{S}_0 \neq 0$, the current derivative must be, $\partial_t \textbf{S} \neq 0$, and vice versa. This inference can be used to explain some superconducting phenomena. This paper can explain two kinds of superconducting currents. a) In the first type of materials, in which only the negative charges can flow, while the positive charges cannot move. When the positive charges are in gradient distribution, the current derivative of the negative charges must not be equal to zero, according to the second-force equilibrium equation, Eq.(14). It states that the negative charges will transfer and may form an electric current, which is the first type of superconducting current. b) In the second type of materials, in which only the positive charges can flow, while the negative charges cannot move. When the negative charges are in gradient distribution, the current derivative of the positive charges will not equal to zero, according to the second-force equilibrium equation, Eq.(14). It means that the positive charges will move and may become the electric current, which is the second type of superconducting current. The two types of superconducting currents (Table \ref{tab:table6}) seem to cover the superconducting currents in the magic-angle graphene superlattices \cite{caoy1,caoy2}.

In case either the electromagnetic strength or gravitational strength is strong enough, the field strength has an influence on the current derivatives and charge gradients to a certain extent, according to the second-force equilibrium equation, Eq.(11). Further the derivatives, $\partial_t \textbf{B}$ , $\partial_t \textbf{E}$ , $\partial_t \textbf{g}$ , and $\partial_t \textbf{b}$ and others are strong enough sometimes, they will exert an impact on the superconducting currents within the conductors, according to the second-force equilibrium equation, Eq.(9).

In terms of some relevant experiment proposals, it is able to validate the second-force equilibrium equation in the laboratories.

\begin{table}[h]
\centering
\caption{Comparison of some equations between the two quaternion spaces, $\mathbb{H}_g$ and $\mathbb{H}_e$ , when the octonion force equals to zero.}
\label{tab:table5}
\begin{tabular}{@{}lllll@{}}
\hline\noalign{\smallskip}
No.    &   formula                   &    equilibrium/continuity equation          &   major physical quantities                               &  space           \\
\noalign{\smallskip}\hline\noalign{\smallskip}
1      &   $\textbf{N}_1^i = 0$      &    force equilibrium equation               &   inertial force, gravity, Lorentz force                  &  $\mathbb{H}_g$  \\
2      &   $N_{10} = 0$              &    fluid continuity equation                &   mass, linear momentum                                   &  $\mathbb{H}_g$  \\
3      &   $\textbf{N}_1 = 0$        &    precession equilibrium equation          &   precessional angular velocity, torque derivative        &  $\mathbb{H}_g$  \\
4      &   $N_{10}^i = 0$            &    torque continuity equation               &   divergence of angular momentum, torque                  &  $\mathbb{H}_g$  \\
5      &   $\textbf{N}_{20} = 0$     &    current continuity equation              &   electric charge, electric current                       &  $\mathbb{H}_e$  \\
6      &   $\textbf{N}_2^i = 0$      &    second-force equilibrium equation        &   current derivative, charge gradient                     &  $\mathbb{H}_e$  \\
7      &   $\textbf{N}_2 = 0$        &    second-precession equilibrium equation   &   current curl, second derivative of magnetic moment      &  $\mathbb{H}_e$  \\
8      &   $\textbf{N}_{20}^i = 0$   &    second-torque continuity equation        &   moment divergence, moment derivative                    &  $\mathbb{H}_e$  \\
\noalign{\smallskip}\hline
\end{tabular}
\end{table}

\section{Experiment proposal}

On the basis of the discharge experiment of lightning rod, it is able to design a suggestion experiment, according to the second-force equilibrium equation. In the experiment, there is not only the electric charge but also the electric current on the conductor of a lightning rod. When the current derivative changes, the charge gradient on the conductor will alter accordingly. Similarly, in case the charge gradient on the conductor varies, the current derivative will modify too.

In the discharge experiment of lightning rod, the wedge shaped end of the lightning rod is positioned to face to the right. Before the lightning is released, the lightning rod is in its state of electrostatic induction. The end of lightning rod will collect comparatively more the electric charges, because the radius of curvature of the end of the lightning rod is quite small. It means that in case the charge gradient, $v_0 \nabla \circ \textbf{S}_0$ , faces toward the right, the current derivative, $\partial_t \textbf{S}$ , on the conductor of a lightning rod must be toward the left, according to the second-force equilibrium equation, Eq.(14).

Referring to some existing experimental schemes (such as, Franklin lightning-rod, early streamer emission lightning-rod, and piezoelectric ceramic discharge device), it is able to verify the second-force equilibrium equation in the experiments. In this proposed experiment, it is necessary to measure the current derivative and charge gradient simultaneously. a) In the first phase, one must determine qualitatively the linear correlation between the current derivative, $\partial_t \textbf{S}$, and the charge gradient, $v_0 \nabla \circ \textbf{S}_0$, in Eq.(14). b) In the second phase, we have to measure quantitatively the accuracy that exists between the current derivative, $\partial_t \textbf{S}$, and the charge gradient, $v_0 \nabla \circ \textbf{S}_0$, in Eq.(14), validating the correlation between the current derivative and charge gradient.

1) Current derivative. At present, there are some experiments that can successfully measure the current derivative, that is, the derivative of the electric current with respect to the time. There are some common methods of measuring the current derivatives, including the differential current coil (or Rogowski coil) \cite{han}, the integral current coil, Faraday magneto-optical effect, and current diverter ring and others. Obviously, some of these experimental schemes can be directly draw on the experience of the suggested experiment of the second-force equilibrium equation, especially the design method of the Rogowski coil sensor.

2) Charge gradient. In the classical field theory, we rarely studied the gradient change of the distribution of the electric charge. Even some scholars think that it is an isolated problem, which seems to possess little research significance. However, in the paper, the charge gradient is an important physical quantity, which is closely related to the current derivative. The charge gradient can exert a significant impact on the current derivative and vice versa. Referring to the existing experimental schemes \cite{sun} relevant to the charge gradient, it is able to measure the charge gradient, in the suggested experiments of the second-force equilibrium equation.

The experiment proposal is helpful to further understand the physical properties of current derivatives. By means of controlling the charge gradient, the amplitude of current derivative can be reduced effectively. This method can enhance the anti-impact ability to resist the high pulse current derivative, diminishing the damage of current derivative to some electronic components, electronic circuits, and high-voltage transmission lines and others.

\begin{table}[h]
\centering
\caption{Comparison of some conditions and current derivatives between the two types of materials, according to the second-force equilibrium equation, Eq.(14).}
\label{tab:table6}
\begin{tabular}{@{}lllll@{}}
\hline\noalign{\smallskip}
material      &   positive charge                  &    negative charge                   &   current derivative         &    conductivity     \\
\noalign{\smallskip}\hline\noalign{\smallskip}
first type    &   fixed, gradient distribution     &    can flow                          &   nonzero                    &    conductor        \\
              &   cannot move                      &    fixed, gradient distribution      &   zero                       &    insulator        \\
second type   &   can flow                         &    fixed, gradient distribution      &   nonzero                    &    conductor        \\
              &   fixed, gradient distribution     &    cannot move                       &   zero                       &    insulator        \\
\noalign{\smallskip}\hline
\end{tabular}
\end{table}

\section{Conclusions and discussions}

The quaternion space $\mathbb{H}_g$ is able to describe the physical properties of gravitational fields, while the 2-quaternion space $\mathbb{H}_e$ is capable of depicting that of electromagnetic fields. The two irrelevant quaternion spaces, $\mathbb{H}_g$ and $\mathbb{H}_e$ , can be considered to be independent of each other. Further they can be combined together to become an octonion space. It means that the octonion space $\mathbb{O}$ can be applied to explore simultaneously the physical properties of gravitational fields and electromagnetic fields, including the octonion field strength, field source, linear momentum, angular momentum, torque, and force and other.

In the octonion space, when the octonion force is equal to zero, it is able to achieve eight isolated equations, such as the force equilibrium equation, fluid continuity equation, current continuity equation, precession equilibrium equation, torque continuity equation, and second-force equilibrium equation. The eight isolated equations are derived from one single equation, and are restricted by each other. If some of the eight equations are correct, so are the rest. It means that the second-force equilibrium equation, $\textbf{N}_2^i = 0$, must be established, in case the force equilibrium equation, fluid continuity equation, and current continuity equation are confirmed.

In the second-force equilibrium equation, the main influencing factors comprise the current derivative, charge gradient, field strength and relative spatial position and others. The current derivative is intensively interrelated to the charge gradient. The current derivative and charge gradient, in the second-force equilibrium equation, are respectively similar to the inertial force and energy gradient, in the force equilibrium equation. By means of some appropriate methods, it is capable of measuring the current derivative and charge gradient simultaneously, verifying the validity of the second-force equilibrium equation in the experiments.

In the engineering applications, according the second-force equilibrium equation, the impact degree of the current derivatives can be reduced by dominating the charge gradients, protecting the electronic components, electronic lines and high-voltage transmission lines and others. Furthermore, the influence of the current derivatives on the electronic facilities can also be lessened by adjusting the field strength and relative spatial positions and others. Obviously, the second-force equilibrium equation has certain practical engineering application values to the electronic products.

In the field theory, the second-force equilibrium equation can be applied to explain the superconducting currents in some materials. When the charge gradient is not equal to zero, the current derivative must not be equal to zero, according to the second-force equilibrium equation, Eq.(14), and vice versa. The existence of current derivatives means that the electric currents may exist in the materials. The paper is able to explain two types of superconducting currents. One is the superconducting currents in the materials where the positive charges cannot be allowed to move. The other is the superconducting currents in the materials which cannot flow with the negative charges.

It is worth mentioning that this paper merely explore a part of major influence factors in the second-force equilibrium equation, but it clearly reveals some close relationships among the current derivative with the charge gradient and others, and the contributions of these physical quantities on the second-force equilibrium equation. The variation of charge gradients may effectively restrain the impact of current derivatives. In the futures studies, we attempt to verify the contribution of charge gradients to the diversification of current derivatives in the experiments, especially the superconducting currents in some materials. It will be helpful to further understand the physical properties of the current derivatives and charge gradients, improving the resistance ability of electronic components and facilities against the impact of current derivatives.

\section*{Acknowledgement}
The author is indebted to the anonymous referees for their valuable comments on the previous manuscripts. This project was supported partially by the National Natural Science Foundation of China under grant number 60677039.

\end{document}